\documentclass[preprint,12pt]{elsarticle}
\usepackage{graphicx}
\usepackage{amsmath}
\usepackage{amssymb} 
\usepackage{xurl}
\usepackage{lineno,hyperref}
\modulolinenumbers[5]
\usepackage{hyperref}% add hypertext capabilities

\journal{Annals of Physics}

\begin{document}

\begin{frontmatter}

\title{Equivalence of a harmonic oscillator to a free particle and 
Eisenhart lift}

\author{Shailesh Dhasmana}
\address{Novosibirsk State University, Novosibirsk 630
090, Russia}
\ead{dx.shailesh@gmail.com}

\author{Abhijit Sen}
\address{Novosibirsk State University, Novosibirsk 630
090, Russia}
\ead{abhijit913@gmail.com}

\author{Zurab K.~Silagadze}
\address{Budker Institute of Nuclear Physics and Novosibirsk State
University, Novosibirsk 630 090, Russia}
\ead{Z.K.Silagadze@inp.nsk.su}

\begin{abstract}
It is widely known in quantum mechanics that solutions of the Schr\"{o}dinger 
equation (SE) for a linear potential are in  one-to-one correspondence with 
the solutions of the free SE. The physical reason for this correspondence 
is Einstein's principle of equivalence. What is usually not so widely known 
is that solutions of the Schr\"{o}dinger equation with harmonic potential can 
also be mapped to the solutions of the free Schr\"{o}dinger equation. The 
physical understanding of this equivalence is not known as precisely as in the
case of the equivalence principle. We present a geometric picture that will 
link both of the above equivalences with one constraint on the Eisenhart 
metric.
\end{abstract}

\end{frontmatter}

\section{Introduction}
There is no clear-cut answer to the question of what geometry is, since ``the 
meaning of the word {\it geometry} changes with time and with the speaker''
\cite{1A}. Classical mechanics was closely related to geometry from the very 
beginning. During the time of Newton and Huygens, many mechanical arguments 
were very geometrical, leading to extensive use of Euclidean  geometry. Then 
came the period when more attention began to be paid to analytical methods, and
Lagrange even boasted that his treatise on analytical mechanics did not 
contain any pictures \cite{1B}. The subsequent development of classical 
mechanics revealed the importance of both analytical and geometric methods. 
The role of geometric ideas and the corresponding intuition became especially 
evident after the works of Poincer\'{e} and Birkhoff. However, the geometries
underlying Hamiltonian mechanics were a new kind of geometry, namely symplectic
geometry and its odd-dimensional cousin contact geometry \cite{1B,1C,1D,1E}  

With the advent of Einstein's general theory of relativity, the role of 
geometry in fundamental physics has increased significantly. General relativity
has two essential features. The first decisive step is the transition from 
three-dimensional Euclidean geometry to four-dimensional spacetime geometry by 
incorporating time as the fourth coordinate\footnote{Interestingly, as early 
as in 1873  P.~L.~Chebyshev, in a letter to J.~J.~Sylvester, wrote ``Take to 
kinematics, it will repay you; it is more fecund than geometry; it adds a 
fourth dimension to space'' \cite{1F,1G}}, and this has already been done by 
Minkowski in the context of special relativity. The second key step is to 
interpret gravity as the curvature of this four-dimensional pseudo-Riemannian
spacetime geometry.

Since Einstein's time, this geometrization of physics has advanced enormously.
In classical physics, all Standard Model fundamental interactions are closely
linked with suitably generalized notions of curvature in geometry. Quantum 
theory adds global (topological) aspects to this local picture of connections
between physics and geometry. As a result, quite diverse, interesting, and 
unexpected geometrical and topological results have emerged from the 
interactions of modern mathematics with fundamental physics \cite{1H}.

Modern differential geometry can be described as the study of (smoothly 
varying) tensor structures on the tangent bundle. In Riemannian geometry, the 
structure is defined in terms of a symmetric, positive definite second rank 
tensor, in contact geometry, in terms of a nondegenerate one-form, and in 
symplectic geometry, in terms of a nondegenerate two-form \cite{1AA}.

If we understand geometrization of nonrelativistic classical mechanics in
a narrower sense as geometrization through Riemannian or Lorentzian 
(pseudo-Riemannian) geometry, we have two options: the ``intrinsic'' approach 
of Cartan and the ``ambient'' approach of Eisenhart \cite{1BB,1CC}. In fact, 
the two approaches are deeply interrelated \cite{1BB,1DD}.
  
Eisenhart \cite{1} made a very important contribution to the geometrization of 
nonrelativistic mechanics. This approach, commonly known as the Eisenhart 
lift, was rediscovered by Duval and coworkers \cite{1DD} in a slightly 
different geometric context of Bargmann structures. Eisenhart showed that 
trajectories of a holonomic conservative dynamical system can be considered as  
null-projections (shadows) of geodesics of some ambient Lorentzian spacetime, 
in amusing analogy with Plato's cave \cite{1CC}.  These ambient spacetimes are
in fact gravitational waves with parallel rays, first considered by  Brinkmann
in  a  different  context \cite{2,2A}. However, Brinkmann did not notice any 
connection with nonrelativistic physics. Be that as it may, historically, 
neither the results of Eisenhart nor Brinkmann actually influenced the 
development of contemporary physics in any way, since they went largely 
unnoticed and then forgotten \cite{1BB}.  

This connection between nonrelativistic and relativistic physics allows one 
to borrow techniques from one field to another, especially the powerful 
geometric tools of Lorentzian geometry. For example, the study of symmetries of 
a nonrelativistic system can be viewed as isometries of the Eisenhart metric. 
Interesting results were obtained in this way \cite{3,3A,3B}.

Duval and his colleagues extended Eisenhart results to nonrelativistic quantum 
mechanics \cite{1DD}, which opened the way for many applications (see, for 
example, \cite{1BB,1CC,K1,K2} and references therein), including applications 
to the nonrelativistic version of the AdS/CFT correspondence \cite{3C,3D}, and 
to the Wheeler-DeWitt equation \cite{S1}.

One of the features of the Eisenhart lift is that the potential term of 
nonrelativistic physics turns into a component of the corresponding Eisenhart 
metric. We can use other coordinates to describe this Brinkman wave and change 
the metric tensor according to the corresponding coordinate transformation. 
When we null-project this new metric back to a nonrelativistic level, we 
get the Schr\"{o}dinger equation with a different potential. Thus, with the 
help of the Eisenhart lift, it is possible to find correspondences between 
the physics of two different physical systems, determined by different 
potentials. 

In this article, we investigate such mappings for the class of potentials 
giving a flat Lorentzian metric after the Eisenhart lift. In section II, we 
briefly review the Eisenhart lift to the extent required for the purposes of 
this study. Next, we will find the flatness condition for the Eisenhart 
metric, which will restrict the choice of the potential. The resulting class 
of potentials and the corresponding equivalencies are explored in the 
subsequent sections.

\section{Eisenhart Lift}
In this section, we'll take a very brief look at the basics of the Eisenhart 
lift. For a more detailed discussion of the material, see \cite{1CC,1DD,1,3,3A,
4,4A,5}, and for a curious analogy to Plato's cave see \cite{5A}.

Consider a general dynamical system described by the Lagrangian
\begin{equation}
L\left(x,\dot{x},t\right) =\frac {1}{2}\,g_{ij}\frac {dx^i}{dt}\frac {dx^j}
{dt}+A_i\left(x,t\right)\frac {dx^i}{dt}-V\left(x,t\right),
\label{eq1}
\end{equation}
where $g_{ij}$ is the so-called  mass matrix, and $A_i$, $V$ are analogous to 
the vector and scalar potentials of the electromagnetic system. To have a
benign nonrelativistic system, we assume that the kinetic term is 
a positive-definite quadratic form in the velocities. In (\ref{eq1}), the time 
$t$ is an evolution parameter that is not part of the configuration-space 
manifold. To transform it into a coordinate on the manifold $(x^i,t)$ and, 
thus, treat it on an equal footing with the spatial coordinates $x^i$, we 
introduce (in the general case,  non-integrable) local time $\tau$ as follows: 
\begin{equation}
m\,d\tau= \Omega\left(x,t\right) dt,
\label{eq2}
\end{equation}
where $m$ is a constant, and its introduction reflects the fact that local 
clocks parameterize trajectories only up to a linear transformation ($m$ 
reflects the arbitrariness in the choice of time units) \cite{5AA}. 

Note that (\ref{eq2}) can be viewed as a generalized Sundman transformation
\cite{5B,5C}\footnote{Sundman introduced the transformation $dt=rd\tau$ in the
context of the three-body problem. See \cite{5D} for a historical account of
Sundman's seminal work on the three-body problem.}. 
Alternatively, in the spirit of the theory of relativity, we can interpret 
$\tau$ as the local time measured  by an arbitrary observer along his/her 
world-line, and then $\Omega$ will serve as the conversion factor to 
Newtonian absolute time $t$. 

As a result the Lagrangian now becomes
\begin{equation}
L=\Omega(x,t) \left\{ \frac{1}{2}\,g_{ij}\frac {dx^i}{d\tau }
\frac {dx^j}{d\tau }+A_i\frac {dx^i}{d\tau }\frac {dt}{d\tau }-
V\frac {dt}{d\tau }\frac {dt}{d\tau }\right\}
\label{eq3}
\end{equation}
The above Lagrangian is obtained by rewriting the action integral in terms of 
the local time $\tau$ and  taking into account the fact that the classical 
action is defined up to a multiplicative constant. We still cannot interpret 
the above Lagrangian as a geodesic Lagrangian for the metric
\begin{equation}
d\bar{s}^2=\Omega \left( x,t\right) \left[ g_{ij}dx^idx^j+2A_idx^idt
-2Vdt^2\right],
 \label{eq4}
\end{equation}
since $\tau$ is not an affine parameter, as it is normalized by $\Omega 
\frac {dt}{d\tau }=m$, and not by the usual normalization condition, which 
is $g_{\mu\nu}\frac{dx^\mu}{d\tau }\frac {dx^\nu}{d\tau }= \mathrm{constant}$. 

To correct this deficiency, we introduce a new coordinate $u$ such that 
\cite{1CC} 
\begin{equation}
du=-Ldt-\frac{M^2}{m}\,d\tau,
\label{eq5}
\end{equation}
where $M$ is some constant. Then
\begin{equation}
\begin{aligned} 
2\,du dt=-\left( g_{ij}dx^idx^j+2A_idx^idt-2Vdt^2\right )-2\,\frac{M^2}{m}
d\tau dt= \\ 
-\left( g_{ij}dx^idx^j+2A_idx^idt-2Vdt^2\right )-\Omega\,\frac{M^2}{m^2}dt^2-
\frac{M^2}{\Omega}d\tau^2, 
\end{aligned}
\label{eq6}
\end{equation}
where we have used (\ref{eq2}) to rewrite the last term in the first line in 
an equivalent form.

After using the above tricks, our final metric becomes 
\begin{eqnarray} &&
ds^2=-M^2d\tau^2=\nonumber \\ && \Omega \left( x,t\right) 
\left\{ g_{ij}dx^{i}dx^{j}+ 2A_{i}dx^{i}dt+2dtdu-2\left(V-\frac {M^2}{2m^2}\,
\Omega\right ) dt^2\right\},
\label{eq7}
\end{eqnarray}
or
\begin{equation}
ds^2=-M^2d\tau^2=g_{\mu\nu}dx^\mu dx^\nu,
\label{eq8}
\end{equation}
and now it is clear that the parameter $\tau$ is normalized in the usual way 
and therefore can be considered as an affine parameter. In (\ref{eq8}) 
$x^\mu=(t,u,x^i)$, $i=1,2,...n$ are coordinates on a $n+2$ dimensional  
Lorentzian manifold.

Thus, the generalized Sundman factor $\Omega$ turns out to be the conformal 
factor relating the ambient metric (\ref{eq7}) to the conformally equivalent 
metric
\begin{equation}
d\tilde{s}^2=g_{ij}dx^{i}dx^{j}+ 2A_{i}(x,t)dx^{i}dt+2dtdu-2U(x,t)dt^2,
\label{eq9}
\end{equation}
with
\begin{equation}
U(x,t)=V(x,t)-\frac {M^2}{2m^2}\,\Omega(x,t).
\label{eq10}
\end{equation}

The ambient metric (\ref{eq7}) was first obtained by Lichnerowicz \cite{6A}, 
and the metric (\ref{eq7}), without the conformal factor $\Omega$, was obtained 
by Eisenhart \cite{1}. Obviously, the conformal factor allows one to 
significantly enlarge the class of the Eisenhart-lifted metrics. 
As we will see later in this article, it is exactly this conformal factor that 
will allow us to find the equivalence of the harmonic oscillator to a free 
particle in the classical and quantum contexts. 

The usefulness of the ambient metric (\ref{eq7}) becomes apparent from the 
following Eisenhart-Lichnerowicz  theorem \cite{1CC}:

{\it Let $\tau$ be an affine parameter such that $\frac{dt}{d\tau}\ne 0$ and
$g_{\mu\nu}\frac{dx^\mu(\tau)}{d\tau}\frac{dx^\nu(\tau)}{d\tau}=-M^2$ on a 
manifold endowed with the metric $$ds^2= \Omega \left( x,t\right) 
\left\{ g_{ij}dx^{i}dx^{j}+2A_{i}dx^{i}dt+2dtdu-2Udt^{2}\right\}.$$
Then the geodesic equation for a curve $x^\mu(\tau)=\left(t(\tau),u(\tau),
x^i(\tau)\right )$, when projected along the direction of $u$ reduces 
to the Euler-Lagrange Equation of the nonrelativistic holonomic  dynamical  
system with the Lagrangian $L=\dfrac {1}{2}\,g_{ij}\frac {dx^i}{dt}
\frac{dx^j}{dt}+A_{i}\frac{dx^i}{dt}-V$, where $V=U+\frac {M^2}{2m^2}\,
\Omega$, with $m=\Omega\, \frac{dt}{d\tau }$.}

\section{Flatness conditions}
For one space dimension, that is, $x^{\mu}=(t,u,x)$, and for $A_i=0$ we get the 
ambient metric as
\begin{equation}
ds^{*2}=\Omega\left(x,t\right) \left\{g(x,t)dx^2+2dtdu-2Udt^2\right\}.
\label{eq11}
\end{equation}
We want the corresponding Riemann curvature tensor to vanish, which will give 
us flat metric conditions for the potential $U(x,t)$ and the conformal factor
$\Omega\left(x,t\right)$.  

For our purposes, we further specialize to the case $g=1$. If the metric
(\ref{eq11}) is flat, then the metric
\begin{equation}
ds^2=dx^2+2dtdu-2Udt^2
\label{eq12}
\end{equation}
is conformally flat. On the other hand, it is well known that in the 
three-dimensional case, a necessary and sufficient condition for the metric to
be conformally flat is the vanishing of the Cotton tensor 
\cite{6B,6C}\footnote{In the four-dimensional case, all conformally flat 
Bargmann manifolds has been determined in \cite{H1} by requiring the vanishing 
of the Weyl tensor. It seems, the problem has not been explored when spacetime 
dimension $d\ge 5$, where the Brinkmann metric may have additional 
components \cite{H2}. These extra components do play a role in the anomaly 
concellation for strings \cite{H3}, so the problem is not only of purely 
academic interest \cite{H4}.}. In terms of the Ricci tensor $R_{\mu\nu}$ and 
the scalar curvature $R$, the three-dimensional Cotton tensor has the form 
\begin{equation}
C_{\mu\nu\lambda}=\nabla_\lambda R_{\mu\nu}- \nabla_\nu R_{\mu\lambda}+\frac{1}{4}\left (
g_{\mu\lambda}\nabla_\nu R-g_{\mu\nu}\nabla_\lambda R\right ).
\label{eq13}
\end{equation}
For the metric (\ref{eq12}), the non-zero Christoffel symbols of the first 
kind  are
\begin{equation}
\Gamma_{ttt}=-\frac{\partial U}{\partial t},\;\;
\Gamma_{ttx}=\Gamma_{txt}=-\frac{\partial U}{\partial x},\;\;
\Gamma_{xtt}=\frac{\partial U}{\partial x}.
\label{eq14}
\end{equation}
Using the inverse metric for  (\ref{eq12}),
\begin{equation}
g^{\mu\nu} =\left (\begin{array}{ccc} 0 & 1 & 0 \\ 1 & 2U & 0 
\\ 0 & 0 & 1
\end{array}\right ),
\label{eq15}
\end{equation}
we find the non-zero Christoffel symbols of the second kind
\begin{equation}
\Gamma^x_{\;tt}=\frac{\partial U}{\partial x},\;\;
\Gamma^u_{\;tx}=\Gamma^u_{\;xt}=-\frac{\partial U}{\partial x},\;\;
\Gamma^u_{\;tt}=-\frac{\partial U}{\partial t}.
\label{eq16}
\end{equation}
To calculate the curvature, it is useful to use Cartan's structure equation
(see, for example, \cite{6D,6E})
\begin{equation}
{\bf R}^\mu_{\;\nu}=d{\bf \Gamma}^\mu_{\;\nu}+{\bf \Gamma}^\mu_{\;\sigma}\wedge 
{\bf \Gamma}^\sigma_{\;\nu},
\label{eq17}
\end{equation}
where (we use bold to indicate differential forms) ${\bf \Gamma}^\mu_{\;\nu}=
\Gamma^\mu_{\;\nu\lambda}\,dx^\lambda$ are connection 1-forms, and ${\bf R}^\mu_{\;\nu}
=\frac{1}{2}\,R^\mu_{\nu\lambda\sigma}dx^\lambda\wedge dx^\sigma$ are curvature 
2-forms. 

From (\ref{eq16}) we find non-zero connection 1-forms
\begin{equation}
{\bf \Gamma}^x_{\;t}=\frac{\partial U}{\partial x}dt,\;\;
{\bf \Gamma}^u_{\;x}=-\frac{\partial U}{\partial x}dt,\;\;
{\bf \Gamma}^u_{\;t}=-\frac{\partial U}{\partial t}dt-
\frac{\partial U}{\partial x}dx,
\label{eq18}
\end{equation}
and using (\ref{eq17}) we find that there are only two non-zero  curvature 
2-forms:
\begin{equation}
{\bf R}^x_{\;t}=\frac{\partial^2 U}{\partial x^2}\,dx\wedge dt,\;\;\;
{\bf R}^u_{\;x}=-\frac{\partial^2 U}{\partial x^2}\,dx\wedge dt.
\label{eq19}
\end{equation}
Therefore, for the metric (\ref{eq12}), non-zero independent components of
the Riemann curvature tensor are
\begin{equation}
R^x_{txt}=\frac{\partial^2 U}{\partial x^2},\;\;\;
R^u_{xxt}=-\frac{\partial^2 U}{\partial x^2}.
\label{eq20}
\end{equation}
As a result, the Ricci tensor has only one non-zero component, and the 
scalar curvature vanishes (since $g^{tt}=0$):
\begin{equation}
R_{tt}=\frac{\partial^2 U}{\partial x^2},\;\;\;
R=g^{\mu\nu}R_{\nu\mu}=0.
\label{eq21}
\end{equation}
We now have all the ingredients for computing the Cotton tensor using 
(\ref{eq13}). The only non-zero components turn out to be
\begin{equation}
C_{ttx}=-C_{txt}=\frac{\partial^3 U}{\partial x^3}.
\label{eq22}
\end{equation}
As we see, a necessary and sufficient condition for the metric (\ref{eq12})
to be conformally flat is
\begin{equation}
\frac{\partial^3 U(x,t)}{\partial x^3}=0.
\label{eq23}
\end{equation}
If two metrics are conformally related $g_{\mu\nu}^*=e^{2\sigma}g_{\mu\nu}$, when 
the corresponding curvature tensors are related as follows \cite{6B}:
\begin{equation}
e^{-2\sigma}R^*_{\mu\nu\lambda\tau}=R_{\mu\nu\lambda\tau}+g_{\mu\tau}\sigma_{\nu\lambda}+
g_{\nu\lambda}\sigma_{\mu\tau}-g_{\mu\lambda}\sigma_{\nu\tau}-g_{\nu\tau}\sigma_{\mu\lambda}+
\left(g_{\mu\tau}g_{\nu\lambda}-g_{\mu\lambda}g_{\nu\tau}\right)\Delta_1\sigma,
\label{eq24}
\end{equation}
where
\begin{eqnarray} &&
\sigma_{\mu\nu}=\nabla_\mu\nabla_\nu\sigma-(\nabla_\mu\sigma)(\nabla_\nu\sigma)=
\nabla_\mu(\partial_\nu\sigma)-(\partial_\mu\sigma)(\partial_\nu\sigma),
\nonumber \\ &&
\Delta_1\sigma=g^{\mu\nu}(\nabla_\mu\sigma)(\nabla_\nu\sigma)=
g^{\mu\nu}(\partial_\mu\sigma)(\partial_\nu\sigma).
\label{eq25}
\end{eqnarray}
Since $R_{tutu}=g_{tt}R^t_{utu}+g_{tu}R^u_{utu}=0$, $g_{uu}=0$ and $\sigma$ does not
depend on $u$, (\ref{eq25}) indicates that when the metric  (\ref{eq11}) is
flat we will have 
\begin{equation}
\Delta_1\sigma=g^{xx}\left (\frac{\partial \sigma}{\partial x}\right )^2=
\left (\frac{\partial \sigma}{\partial x}\right )^2=0.
\label{eq26}
\end{equation}
Therefore, when the Eisenhart metric (\ref{eq11}) is flat, the local time  
(\ref{eq2}) is integrable, since $\sigma$ (and thus $\Omega$) is independent 
of $x$.

The conformal transformation law for the three-dimensional Ricci tensor is  
\cite{6B,6F,6G} (note that in \cite{6B} the Ricci tensor is defined with 
the opposite sign)
\begin{equation}
R^*_{\mu\nu}=R_{\mu\nu}-\sigma_{\mu\nu}-g_{\mu\nu}\left (\Delta_2\sigma+\Delta_1\sigma
\right ),\;\;\;\Delta_2\sigma=g^{\mu\nu}\nabla_\mu\nabla_\nu\sigma=
g^{\mu\nu}\nabla_\mu\left(\partial_\nu\sigma\right).
\label{eq27}
\end{equation}
As we have seen, the only non-zero component of the Ricci tensor is $R_{tt}$.
Since $\sigma$ does not depend on $x$, (\ref{eq15}) indicates that both
$\Delta_2\sigma$ and $\Delta_1\sigma$ are zero, and as a flatness condition
we get $R_{tt}=\sigma_{tt}=\partial_t \partial_t\sigma-(\partial_t\sigma)^2$ 
(we have taken into account that $\nabla_t(\partial_t\sigma)=\partial_t 
\partial_t\sigma -\Gamma^x_{tt}\,\partial_x\sigma-\Gamma^u_{tt}\,\partial_u
\sigma=\partial_t \partial_t\sigma$). 

In terms of $\Omega=e^{2\sigma}$ the flatness conditions takes the form
\begin{equation}
\frac {\partial \Omega }{\partial x}=0,\;\;\;
\frac {\partial^2 U}{\partial x^2}=\frac {1}{\Omega^2}
\left\{ \frac {\Omega}{2}\,\frac {\partial^2\Omega }{\partial t^2}-
\frac {3}{4}\left(\frac {\partial \Omega }{\partial t}\right)^2\right\}. 
\label{eq28}
\end{equation}
Note that (\ref{eq23}) follows from (\ref{eq28}), so (\ref{eq28}) is
a necessary and sufficient condition the metric (\ref{eq11}) (with $g=1$) 
to be flat. In addition, as $\Omega$ is $x$-independent, it follows from 
(\ref{eq10}) that $\frac {\partial^2 U}{\partial x^2}=\frac {\partial^2 V}
{\partial x^2}$, and the conditions (\ref{eq28}) do not depend on $M$. 
In particular, we can take $M=0$, which corresponds to null-geodesics in 
the ambient spacetime. Accordingly, in what follows we replace $U$ by $V$.

Thus, under our assumptions, we finally obtain a class of potentials for which 
the Eisenhart metric (\ref{eq11}) is flat, or, in other words, the metric can 
be brought to the standard Minkowski form  by means of a suitable coordinate 
transformation. It is somewhat surprising that only spatially linear, or 
quadratic potentials are allowed by the flatness condition, while there is 
still huge scope for temporal dependence. Nevertheless, the Eisenhart metric
(\ref{eq11}) allows to analyze interesting cases such as harmonic 
oscillator (both time-dependent and time-independent), linear potential etc.  

For further analysis, it is useful to introduce a function $\Phi(t)$ such
that $\Omega=\frac{d\Phi}{dt}$ and rewrite the second equation in 
(\ref{eq28}) in the form
\begin{equation}
\frac {\partial^2 V}{\partial x^2}=\frac{1}{2}\,S_t(\Phi),\;\;\;
S_t(\Phi)=\frac {1}{\frac {d \Phi}{d t}}\frac {d ^3\Phi}
{d t^3}-\frac {3}{2}\frac {1}{\left( \frac {d \Phi}{d t}
\right)^2}\left( \frac {d ^2 \Phi}{d t^2}\right)^2.
\label{eq29}
\end{equation}
Here $S_t(\Phi)$ is the so-called Schwarzian derivative of the function
$\Phi(\tau)$. Despite its rather complicated form, the Schwarzian derivative
is ubiquitous and tends to appear in many seemingly unrelated fields of 
mathematics \cite{6AA,6BB,6CC}.

Now, since we have a metric that is flat under condition (\ref{eq29}), we will 
try to find a coordinate transformation that brings this metric 
\begin{equation}
ds^{*2}=\Omega\left(t\right) \left\{dx^2+2dtdu-2Vdt^2\right\}.
\label{eq30}
\end{equation}
to the standard Minkowski form.  It can be checked rather easily that the 
transformation 
$(t,u,x)\rightarrow (\tau ,v,\xi)$, with
\begin{equation}
x=f_1\left( \xi ,\tau \right), \;\;\; t=f_2\left(\xi ,\tau \right), \;\;
u=v+f_3\left( \xi ,\tau \right),
\label{eq31}
\end{equation}
brings the metric (\ref{eq30}) in the Minkowski form $ds^{*2}=d\xi ^2+
2d\tau dv$ if
\begin{eqnarray} &&
\frac{\partial f_1}{\partial \xi }=\frac{1}{\sqrt {\Omega(t) }},\;
\frac{d f_2}{d \tau }=\frac{1}{\Omega(t) }, \;
\frac {\partial f_{2}}{\partial \xi}=0, \nonumber \\ &&
\left(\frac {\partial f_1}
{\partial \tau }\right) ^2+\frac{2}{\Omega(t) }
\,\frac {\partial f_3}{\partial \tau } -\frac {2V}
{\Omega(t) ^2}=0,\; \sqrt{\Omega(t) }\;\frac {\partial f_1}
{\partial \tau } +\frac{\partial f_3}{\partial \xi } =0.
\label{eq32}
\end{eqnarray}
It is clear from the second and third equations in (\ref{eq32}) that 
$t=f_2(\tau)$ is a function of $\tau$ only. This function can be found
by inverting the relation $\tau =\int \Omega(t)dt$ and then the system 
(\ref{eq32}) has a solution
\begin{equation}
x=f_1\left( \xi,\tau\right) =\frac {\xi }{\sqrt {\Omega(t) }}+
h\left( \tau \right),\;\;\;
f_3\left( \xi,\tau \right) =\frac {\xi ^2}{4\Omega(t) } 
\frac {d \Omega(t)}{d \tau } -\sqrt{\Omega(t) }\;
\frac {d h}{d \tau }\,\xi +p\left( \tau \right),
\label{eq33}
\end{equation}
where the functions $p(\tau)$ and $h(\tau)$ are constrained by the fourth
equation in (\ref{eq32}). In particular, for this equation to be valid for 
all values of $\tau$ and $\xi$, the functions $p(\tau)$ and $h(\tau)$ must
satisfy the equations
\begin{equation}
\frac{dp}{d\tau}=-\frac{\Omega(t)}{2}\left(\frac{dh}{d\tau}\right)^2+
\frac{S_t[\Phi]\,h^2}{4\Omega(t)}+\frac{B(t)\,h}{\Omega(t)}+\frac{C(t)}
{\Omega(t)},
\label{eq34}
\end{equation}
and
\begin{equation}
\frac{d^2 h}{d \tau^2}+\frac{1}{\Omega(t)}\,\frac{d\Omega(t)}{d \tau}\,
\frac{d h}{d \tau}+\frac{S_t[\Phi]\, h}{2\Omega^2(t)}=-\frac{B(t)}
{\Omega^2(t)}.
\label{eq35}
\end{equation}
To derive the above equations, we used the most general form of the 
potential, compatible with (\ref{eq29}): 
\begin{equation}
V(x,t)=\frac{1}{4}S_t(\Phi)x^2+B(t)x+C(t).
\label{eq36}
\end{equation}
Here $B(t)$ and $C(t)$ are arbitrary functions of time. 

Equations (\ref{eq33})-(\ref{eq35}) constitute the required coordinate 
transformation for the potential (\ref{eq36}). Different choices of the 
potential $V(x,t)$ correspond to different forms of $S_t(\Phi(t))$, from 
which we get $\Phi(t)$ and hence $\Omega(t)$. Then, in principle, we can 
find the function $t(\tau)$, solve the differential equation (\ref{eq35}) 
for $h(\tau)$ and then determine $p(\tau)$ from (\ref{eq34}). Finally, we get 
the required coordinate transformation using (\ref{eq33}). 

Let us see how the nonrelativistic Lagrangian $L_V=\frac{1}{2}\left (
\frac{dx}{dt}\right )^2-V(x,t) $ is related to the free Lagrangian $L_{free}=
\frac{1}{2}\left (\frac{d\xi}{d\tau}\right )^2$. We have 
$$\frac{dx}{dt}=\Omega(t)\left [\frac{\partial f_1}{\partial \xi}\,
\frac{d\xi}{d\tau}+\frac{\partial f_1}{\partial \tau}\right ]=
\sqrt{\Omega(t)}\left [\frac{d\xi}{d\tau}+\sqrt{\Omega(t)}\,\frac{\partial f_1}
{\partial \tau}\right ],$$ 
and using relations from (\ref{eq32}), we get
$$\begin{aligned}
& \frac{1}{2}\left (\frac{dx}{dt}\right )^2=\Omega(t)\left [\frac{1}{2}
\left (\frac{d\xi}{d\tau}\right )^2-\frac{\partial f_3}{\partial \tau}-
\frac{d\xi}{d\tau}\,\frac{\partial f_3}{\partial \xi}\right ]+V(x,t)= \\
& \Omega(t)\left [\frac{1}{2}\left (\frac{d\xi}{d\tau}\right )^2-
\frac{df_3}{d\tau}\right ]+V(x,t).\end{aligned}$$ 
Therefore, $L_V=\Omega(t)\left[L_{free}-\frac{df_3}{d\tau}\right]$ and we see 
that the Lagrangian does not remain invariant under transformations 
(\ref{eq33}). However, the actions $S_V=\int L_V dt$ and $S_{free}= \int L_{free}
\,d\tau$ are equivalent. Indeed,
\begin{equation}
S_V=\int\limits_{t_i}^{t_f} L_V dt =\int\limits_{t_i}^{t_f}{\Omega(t)\left[L_{free}-
\frac{df_3}{d\tau}\right]}dt,
\label{eq37}
\end{equation}
and using $d\tau=\Omega(t)dt$ from (\ref{eq32}), we get
\begin{equation}
S_V=\int\limits_{\tau_i}^{\tau_f}{\left[L_{free}-\frac{df_3}{d\tau}\right]}d\tau=
S_{free}-\left [f_3\left (\xi(\tau_f),\tau_f\right )-f_3\left (\xi(\tau_i),
\tau_i\right )\right ].
\label{eq38}
\end{equation}
The boundary terms in square brackets do not affect Euler-Lagrange 
equations of motion.

The quantity $e^{iS}$, which is central in the path-integral formulation of 
quantum mechanics, when the coordinates change according to (\ref{eq33}), 
transforms as $e^{iS_V}\rightarrow e^{-if_3}e^{iS_{free}}$. Thus, $f_3$ is a phase 
factor. Although the transformation of the Eisenhart metric is purely 
classical, it is surprising that it can generate a quantum mechanical phase 
factor. This hints at the possibility that the Eisenhart lift can also 
incorporate quantum transformations, with the auxiliary coordinate $u$ 
acting as a phase factor. In the next section, we will find out exactly how 
the wave functions are transformed when the coordinates change according to 
(\ref{eq33}), and we will show that the function $f_3$ is indeed a phase 
factor accompanying the transformation of the wave function.
 
\section{Schr\"{o}dinger Equation and Eisenhart lift}
It is known that the Schr\"{o}dinger equation can be derived as a null 
dimensional reduction of the Klein-Gordon equation \cite{1CC,1DD}. This 
connection has interesting applications because the Eisenhart metric 
(\ref{eq7}) (called Platonic waves in \cite{1CC}) covers a large class of 
spacetimes of physical  and mathematical importance \cite{1CC}. After the
null dimensional reduction, various Eisenhart metrics will correspond 
to different potentials in the Schr\"{o}dinger equation. For example, a class 
of spacetimes, called AdS-pp-waves, can be obtained using 
$\Omega =\frac {1}{x^{2}}$ and arbitrary scalar potential $V(x,t)$ in 
(\ref{eq30}). Further specialization $V\sim -x^{2\left( 1-Z\right) },\;Z\ge 1$ gives 
us the so called Schr\"{o}dinger spacetime. Anti-de Sitter spacetime 
corresponds to $Z=1$. 

Interestingly, the Schr\"{o}dinger spacetimes have applications to 
nonrelativistic holography \cite{21}. Another important observation is that 
in \cite{22} Penrose proved that near a null geodesic, every Lorentzian
spacetime looks like a pp-wave with metric which is exactly the 
Eisenhart metric in two spatial dimensions with $\Omega=1$ and 
$V=a(t)(x^2-y^2)+2b(t)xy$ for some $a(t)$, $b(t)$. This result is known as the 
Penrose limit in the physical literature \cite{23}. 

Since Eisenhart metric covers such a wide class of important manifolds, the 
study of the quantum field theory in these Eisenhart spacetimes is expected 
to be significantly simplified by converting the problem to the corresponding 
Schr\"{o}dinger equation. On the contrary, the understanding of the 
corresponding Schr\"{o}dinger equation can help in the study of Eisenhart 
spacetimes. Below this correspondence between the Eisenhart metric and the 
Schr\"{o}dinger equation is illustrated in the context of the flat Eisenhart 
metric (\ref{eq30}).

\begin{figure}[h]
    \centering
    \includegraphics[width=0.5\textwidth]{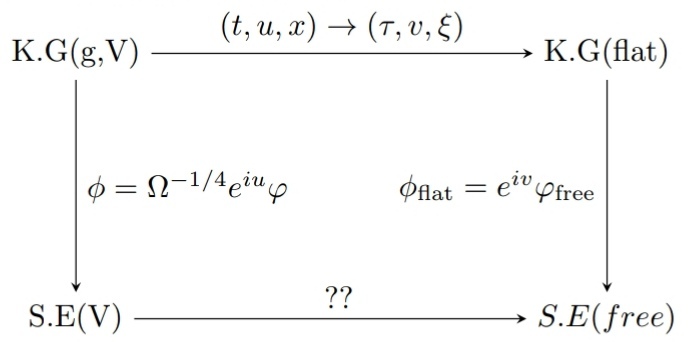}
    \caption{Schematic diagram showing how the transformation of the  
wave function is obtained.}
    \label{fig:1}
\end{figure}

Let $x^{\mu}=(t,u,x)$, and consider the Klien-Gordon equation of a massless 
scalar field $\phi$ in the 3-dimensional Eisenhart metric (\ref{eq30})
\begin{eqnarray} &&
\square \phi =\frac {1}{\sqrt {-g}}\,\partial _{\mu }\left( \sqrt {-g}\,
g^{\mu\nu}\partial _{\nu}\phi \right) =0, \nonumber \\ &&
g_{\mu\nu}=\begin{bmatrix} 
-2\Omega V & \Omega & 0 \\ \Omega & 0 & 0 \\ 0 & 0 &  \Omega \end{bmatrix},
\;\;\;g^{\mu\nu}=\begin{bmatrix} 0 & \Omega^{-1} & 0 \\ \Omega^{-1} & 
2\Omega^{-1} V & 0 \\ 0 & 0 &  \Omega^{-1} \end{bmatrix}.
\label{eq39}
\end{eqnarray}
In the explicit form the equation is 
\begin{equation}
\sqrt{\Omega}\,\frac {\partial ^{2}\phi }{\partial x^{2}}+2\sqrt {\Omega}\,V 
\frac {\partial ^{2}\phi }{\partial u^{2}}+\frac {\partial \sqrt {\Omega}}
{\partial t}\,\frac {\partial \phi }{\partial u}+2\sqrt {\Omega}\,
\frac {\partial ^{2}\phi }{\partial t\partial u}=0,
\label{eq40}
\end{equation}
which, after the field transformation
\begin{equation}
\phi(t,x) =\Omega ^{-1/4}e^{iu}\varphi \left( t,x\right),
\label{eq41}
\end{equation} 
is reduced to the Schr\"{o}dinger equation
\begin{equation}
i\frac {\partial \varphi }{\partial t}=-\frac {1}{2}\frac {\partial ^{2}
\varphi }{\partial x^{2}}+V\varphi.
\label{eq42}
\end{equation}
An interesting consequence of the transformation of the Klein-Gordon problem 
into the Schr\"{o}dinger problem is that it can be used to find a map 
between the solution of the Schr\"{o}dinger equation in a given potential and 
a free solution using the coordinate transformation and the covariance of the 
scalar field under this transformation. The conditions for the existence of 
such a transformation of coordinates will constrain the class of potentials 
that can be mapped onto the problem of free particles by this method. We have 
already found such a class of potentials in (\ref{eq29}) and the corresponding 
transformation of coordinates that leads to a free system in (\ref{eq33}).
For clarity, see Fig.\ref{fig:1}, which shows the various transformations. 
The relationship between the wave functions is obtained by requiring the 
diagram to be commutative. This commutativity is possible because of the 
covariance of the scalar field under the coordinate transformation.
Let us find explicitly the mapping between the wave functions. Using the 
covariance of the scalar field, $\phi(t,u,x)=\phi_{flat}(\tau,v,\xi)$ under 
the transformation $(t,u,x)\rightarrow (\tau,v,\xi)$, and the relation 
(\ref{eq41}), we obtain $\Omega^{-1/4}e^{iu}\varphi_{V}=e^{iv}\varphi_{free}$ and,
therefore,
 \begin{equation}
\varphi_{V}(t,x)= \Omega ^{1/4}e^{-i(u-v)}\varphi_{free}(\tau,\xi).
\label{eq43}
\end{equation}
But $u-v=f_3$, and using (\ref{eq33}) and (\ref{eq34}), we get finally the 
wave function transformation
\begin{equation} 
\varphi_{V}(t,x)=\Omega ^{1/4}(t(\tau))
e^{-if_3(\tau,\xi)}\,\varphi_{free}(\tau,\xi),
\label{eq44}
\end{equation}
where
\begin{eqnarray} &&  
f_3(\tau,\xi)=\frac {\xi ^2}
{4\Omega(t(\tau)) }\frac {d \Omega(t(\tau)) }{d \tau } -
\sqrt{\Omega(t(\tau)) }\,\frac {d h}{d \tau }\,\xi 
\nonumber \\ &&
+\int{\left[\frac{\Omega(\tau)}{2}\left(\frac{dh(\tau)}{d\tau}\right)^2
+\frac{S_{\tau}[\Phi(\tau)]h^2(\tau)}{4\Omega(\tau)}+\frac{B(\tau)h(\tau)+
C(\tau)}{\Omega(\tau)}\right]}d\tau. \quad
\label{eq44A}
\end{eqnarray}
This is the required mapping we were looking for. Here $\varphi_V(x,t)$ and 
$\varphi_{free}$ are solutions of the Schrodinger equation with potential 
$V(x,t)$ (under the constraint (\ref{eq29})) and with zero potential, 
respectively. 

It is interesting to note that the transformation of the Eisenhart metric 
immediately gives us both the classical coordinate transformation and the 
quantum wave function transformation. The function $f_3$, as anticipated at 
the end of the previous section, appears as a phase in the transformation of 
the wave function. 

In the following sections, we will explicitly find the most general maps 
between specific potentials and zero potential. In particular, we consider 
harmonic potential in section V and linear potential in section VI. We will 
deal with both time-independent and time-dependent cases and show that the 
previous results found in the literature are obtained as trivial sub-cases of 
our general results.

\section{Harmonic oscillator to a free particle transformation}
The relationship between quantum particles in zero and harmonic potentials 
appears to have been identified first in the field of optics 
\cite{23A,23B,23C}. The full mathematical equivalence of these problems was
established by Niederer using group theoretical arguments \cite{23D}. 
In \cite{8} Takagi explicitly showed how a quantum particle in harmonic 
potential can be mapped into a free particle. A fairly general result of this 
kind was also obtained by Arnold in the analysis of differential equations 
\cite{1C}. 

In this section, we show that these results are all special cases of a 
very general coordinate transformation. This will be done using the Eisenhart 
lift method as discussed in the previous sections. It is interesting to note 
that in the previous works of Takagi and Arnold, the  mapping between a 
particle in harmonic potential and a free particle appears to be an ingenious 
guess. While in our case we are trying to classify all the potentials that 
will give us the flat Eisenhart metric, and by investigating this flatness 
condition, we find new transformations as a by-product and at the same time 
reproduce all the previous results already existing in the physical 
literature. In this respect, the Eisenhart lift method is more elegant.
 
As we have already mentioned, the most general potential compatible with the 
flatness condition is (\ref{eq36}). This covers the cases of time-dependent 
and time-independent harmonic potentials and linear potential\footnote{It is 
interesting that only for such potentials does quantum dynamics in phase 
space demonstrate Liouvillian behavior at all times \cite{O1,O2}.}.

The time-dependent harmonic potential, $V(x,t)=\frac {1}{2}\omega^2 (t)x^{2}$, 
corresponds to  $B(t)=C(t)=0$ and
\begin{equation}
S_t(\Phi)=2\omega^2 (t).
\label{eq45}
\end{equation}
According to the general theory of Schwarzian derivative \cite{6DD}, 
(\ref{eq45}) means that,
$\Phi(t)=\frac{u_1(t)}{u_2(t)},$
where $u_1$ and $u_2$ are two linearly independent solutions of
\begin{equation}
\frac{d^2u(t)}{dt^2}+\omega^2 (t)u(t)=0.
\label{eq46}
\end{equation}
Then $\Omega(t)=\frac{d\Phi}{dt}=\frac{(u_2\frac{du_1}{dt}-u_1\frac{du_2}{dt})}
{u_2^2}$. Let us choose the normalizations of $u_1$ and $u_2$ so that the 
Wronskian $W(u_1, u_2)=u_2\frac{du_1}{dt}-u_1\frac{du_2}{dt}=1$. In this case 
$\Omega(t)=\frac{1}{u_2^2(t)}$ and the equation for $h(\tau)$ takes the form
\begin{equation}
\frac{d^2 h(\tau)}{d \tau^2}-\frac{2}{u_2(t(\tau))}
\frac{d u_2(t(\tau))}{d \tau}\,\frac{d h(\tau)}
{d \tau}+\omega^2(t(\tau))u_2^4(t(\tau))h(\tau)=0.
\label{eq47}
\end{equation}
For a given functional form of the frequency $\omega(t)$, we can solve 
(\ref{eq46}) and find $u_2$, which in principle will allow us to solve 
(\ref{eq47}) and find $h(\tau)$ and $p(\tau)$. Since $\tau=\int \Omega(t)dt=
\Phi(t)$, from (\ref{eq31}) and (\ref{eq33}) we get the following coordinate
transformations:
\begin{eqnarray} &&
\tau=\frac{u_1(t)}{u_2(t)},\;\;\;
\xi=\frac{x-h(\tau(t))}{u_2(t)},\nonumber \\ &&
v=u+\frac {[x-h(\tau(t))]^2}{2u_2(t)}\frac{du_2(t)}{dt}+\left[x-h(\tau(t))
\right]\frac{dh(\tau(t))}{dt}-p(\tau(t)). \quad
\label{eq48}
\end{eqnarray}
Whereas, according to (\ref{eq44}), the wave function is transformed as 
follows:   
\begin{equation}
\varphi_{free}(\tau,\xi)= \sqrt{u_2(t)}\,e^{i\left[-\frac {[x-h(\tau(t))]^2}
{2u_2(t)}\frac{du_2(t)}{dt}-\left[x-h(\tau(t))\right]\frac{dh(\tau(t))}{dt}
+p(\tau(t))\right]}\varphi_{H.O}(t,x), 
\label{eq49}
\end{equation}
Equations (\ref{eq48}) and (\ref{eq49}) define the most general quantum 
transformation between a harmonic oscillator and a free particle. The trivial 
solution $h(\tau)=0$ of (\ref{eq47}) corresponds to $\tau=\frac{u_1(t)}
{u_2(t)}$, $\xi=\frac{x}{u_2(t)}$, which is exactly the Arnold 
transformation \cite{1C}, and together with the transformation of the 
wave function
$$\varphi_{free}(\tau,\xi)= \sqrt{u_2(t)}\,e^{-i\left[\frac {x^2}{2u_2}
\frac{du_2}{dt}\right]}\varphi_{H.O}(t,x)$$ 
they constitute the so-called "Quantum Arnold transformation" \cite{26}. 

In the case of a time-independent Harmonic potential, $\omega(t)=\omega_0$, 
two linearly independent solutions of (\ref{eq46}) will be $u_1(t)=
\frac{1}{\omega_0}sin(\omega_0 t)$ and $u_2(t)=cos(\omega_0 t)$. Our choice of 
$u_2$ is such that at $t=0$, $\Omega$ is well defined, and the normalization
of $u_1$ is chosen in such a way as to make the Wronskian $W(u_1, u_2)=1$. 
Then $\Omega(t)=\frac{1}{u_2^2(t)}=\frac{1}{\cos^2{\omega_0 t}}=
\sec^2(\omega_0 t)$ and   $\tau=\frac{\tan(\omega_0 t)}{\omega_0}$. In terms of 
$\tau$, we get $\Omega(\tau)=(1+\omega_0^{2}\tau ^{2})$ and 
$u_2(\tau)=\frac{1}{\sqrt {1+\omega_0^{2}\tau ^{2}}}$. Substituting $u_2(\tau)$ 
in (\ref{eq47}), we get 
\begin{eqnarray} &&
\frac{d ^2h(\tau)}{d \tau^2}+\frac{2\omega_0^{2}
\tau}{1+\omega_0^{2}\tau ^{2}}\,\frac{d h(\tau)}{d \tau}+\frac{\omega_0^2}
{(1+\omega_0^{2}\tau ^{2})^2}\,h(\tau)= \nonumber \\ &&
\frac{1}{\sqrt{1+\omega_0^{2}\tau ^{2}}}\,
\frac{d^2}{d\tau^2}\left[h(\tau)\,\sqrt{1+\omega_0^{2}\tau ^{2}}\right]=0.
\label{eq50}
\end{eqnarray}
Therefore, the general form of the function  $h(\tau)$ is
\begin{equation}
h(\tau)=\frac{c_1+c_2\,\omega_0\tau}{\sqrt{1+\omega_0^{2}\tau ^{2}}},
\label{eq51}
\end{equation} 
where $c_1$ and $c_2$ arbitrary constants.

We can find $p(\tau)$ from (\ref{eq34}):
\begin{eqnarray} &&
p(\tau)=\int\left[-\frac{(1+\omega_0^{2}\tau ^{2})}{2}
\left(\frac{dh(\tau)}{d\tau}\right)^2+\frac{\omega_0^2h^2(\tau)}
{2(1+\omega_0^{2}\tau ^{2})}\right]d\tau= \nonumber \\ &&
\frac{(c_1^2-c_2^2)\omega_0^2\tau-2c_1c_2\omega_0}{2(1+\omega_0^2\tau^2)}+c_3,
\label{eq52}
\end{eqnarray}
Where $c_3$ is an integration constant. Therefore, ultimately the 
transformation of coordinates takes the form 
\begin{eqnarray} &&
x=\frac {\xi+c_2\,\omega_0\tau+c_1}{\sqrt {1+\omega_0^{2}\tau ^{2}}},\;\;\;
t=f_{2}\left( \tau \right) =\frac {1}{\omega }\tan ^{-1}\left( \omega \tau 
\right),\;\;\; u=v+f_3(\xi, \tau)= \nonumber \\ &&
v+\frac {\omega_0^{2}\xi ^{2}\tau }{2(1+\omega_0 ^{2}
\tau ^{2})}-\frac{(c_2-c_1\,\omega_0\tau)\,\omega_0}{1+\omega_0^{2}\tau ^{2}}\,\xi 
+\frac{(c_1^2-c_2^2)\omega_0^2\tau-2c_1c_2\,\omega_0}{2(1+\omega_0^2\tau^2)}+c_3,
\qquad 
\label{eq53}
\end{eqnarray}
and the transformation law of the wave function becomes 
\begin{eqnarray} &&
\varphi_{H.O}(t,x)= (1+\omega_0^{2}\tau ^{2})^{1/4} \times \nonumber \\ &&
e^{-i\left(\frac {\omega_0^{2}\xi ^{2}\tau}{2(1+\omega_0 ^{2}\tau^{2})}-
\frac{(c_2-c_1\omega_0\tau)\omega_0}
{(1+\omega_0^{2}\tau ^{2})}\xi +\frac{(c_1^2-c_2^2)\omega_0^2\tau-2c_1c_2\omega_0}
{2(1+\omega_0^2\tau^2)}+c_3\right)}\varphi_{free}(\tau,\xi)
\label{eq54}
\end{eqnarray} 
Equations (\ref{eq53}) and (\ref{eq54}) represent the most general and 
complete class of quantum transformations between a time-independent harmonic 
oscillator and free particle.
 
An important observation is that the transformations (\ref{eq54}) are 
non-unitary due to the presence of the time dependent factor 
$(1+\omega_0^2\tau^2)^{1/4}$. Thus, these transformations cannot represent the 
principle of equivalence, in contrast to the case of linear potential 
(gravity), considered below. It is impossible to reconstruct the entire 
temporal history of a harmonic oscillator from the temporal history of 
a free particle using only one transformation (\ref{eq53})  (or (\ref{eq48}) 
in the case of time-dependent oscillator). This situation is analogous to 
the situation in geometry then a manifold cannot be covered by a single 
patch of coordinates \cite{8}.

In the limiting case $c_1=0=c_2$, which implies $h(\tau)=0$  and $p(\tau)=0$, 
one recovers the solution already known in the literature as the Niederer 
transformation \cite{23D,25,27}\footnote{A generalization of the Niederer 
transformation for oscillators with time-dependent frequency has been studied 
very recently in \cite{27A}.}. In fact, all the previous results relating 
one-dimensional harmonic potential to zero potential are special cases 
of the most general transformation (\ref{eq48}).

\section{Mapping a linear potential to a free particle}
Having considered in detail the case of a harmonic oscillator in the previous 
section, we now consider the case of a linear potential. A linear  potential
$V=g(t)\,x$ can be associated with a weak gravitational field. A beautiful 
neutron interference experiment \cite{29} revealed quantum effects in 
such fields. Mathematically, these quantum effects arise from the phase 
transformation of the wave function caused by the transformation of 
coordinates in the Schr\"{o}dinger equation during the transition from an 
inertial coordinate system to an accelerated one \cite{29A}. The same effect 
was observed when setting the entire neutron interferometer into harmonic 
oscillations and measuring the phase shift as a function of the apparatus 
maximal acceleration, which demonstrates the validity of the equivalence 
principle in the nonrelativistic quantum regime \cite{29B}. Sakurai's book 
\cite{28} contains a brief pedagogical account of neutron interference 
phenomena in the presence of a weak gravitational potential. 

In this section, we will discuss the equivalence between a linear potential 
and a free particle using the Eisenhart lift and find out explicitly 
the most general  coordinate transformations together with the corresponding 
quantum phase transformations. It is shown that the transformation 
representing the equivalence principle is a special case of these general 
transformations. We report a new class of non-unitary transformations 
previously unknown to the authors. We consider both time-independent and 
time-dependent cases. 

Time-dependent linear potential, $V(x,t)=B(t)\,x$, corresponds to $C(t)=0$
and $S_t(\Phi)=0$ in (\ref{eq36}). From the theory of the Schwarzian 
derivative, we know that the general solution of the equation  $S_t(\Phi)=0$ 
is given by the M\"{o}bius transformation $\Phi(t)=\frac{at+b}{ct+d}$, 
$k=ad-bc\ne 0$. First consider the case $c\ne 0$. Then $\Phi(t)=\frac{a}{c}-
\frac{k}{c(ct+d)}$, and $\Omega(t)=\frac{d\Phi}{dt}=\frac{k}{(ct+d)^2}$, 
$\tau=\int \Omega(t)dt=\frac{-k}{c(ct+d)}$. Therefore, $\Omega(\tau)=
\frac{c^2\tau^2}{k}$. If we substitute this into (\ref{eq35}), we get
\begin{equation}
\frac{d^2h(\tau)}{d \tau^2}+\frac{2}{\tau}\frac{d h(\tau)}{d \tau}=
\frac{1}{\tau^2}\,\frac{d}{d\tau}\left (\tau^2\,\frac{dh(\tau)}{d\tau}\right )=
-\frac{B(t(\tau))\,k^2}{c^4\tau^4}.
\label{eq55}
\end{equation} 
The general solution of this differential equation is 
\begin{equation}
h(\tau)=-\frac{k^2}{c^4}\int\frac{1}{\tau^2}\left[\int\frac{B(t(\tau))}
{\tau^2}d\tau\right]d\tau-\frac{c_1}{\tau}+c_2,
\label{eq56}
\end{equation}
where $c_1$, $c_2$ are arbitrary constants. Then $p(\tau)$ is determined 
from (\ref{eq34}):
\begin{eqnarray} &&
p(\tau)=\int\left[\frac{-c^2\tau^2}{2k}\left(-\frac{k^2}{c^4\tau^2}\int
\frac{B(t(\tau))}{\tau^2}d\tau+\frac{c_1}{\tau^2}\right)^2+ \right .
\nonumber \\ && \left .
\frac{B(t(\tau))k}{c^2\tau^2}\left(-\frac{k^2}{c^4}\int\frac{1}{\tau^2}
\left[\int\frac{B(t(\tau))}{\tau^2}d\tau\right]d\tau-\frac{c_1}{\tau}+
c_2\right)\right]d\tau.
\label{eq57}
\end{eqnarray}
Correspondingly, according to (\ref{eq33}), the coordinate transformation 
becomes 
\begin{equation}
\begin{aligned}
& t=\frac{-k}{c^2\tau}-\frac{d}{c},\;\;\;
x=\frac{\sqrt{k}\,\xi}{c\tau}-\frac{k^2}{c^4}\int\frac{1}{\tau^2}\left[\int
\frac{B(t(\tau))}{\tau^2}d\tau\right]d\tau-\frac{c_1}{\tau}+c_2,\\
& f_3=\frac{\xi^2}{2\tau}-\frac{c\tau\,\xi}{\sqrt{k}}\left(-\frac{k^2}{c^4\tau^2}
\int\frac{B(t(\tau))}{\tau^2}d\tau+\frac{c_1}{\tau^2}\right)+p(\tau),
\end{aligned}
\label{eq58}
\end{equation}
and the wave function transforms as
\begin{equation}
\varphi_{grv}(t,x)=\frac{\sqrt{c\,\tau}}{k^{1/4}}\, 
e^{-i\left[\frac{\xi^2}{2\tau}-\frac{c\tau\xi}{\sqrt{k}}\left(-\frac{k^2}
{c^4\tau^2}\int\frac{B(t(\tau))}
{\tau^2}d\tau+\frac{c_1}{\tau^2}\right)+p(\tau)\right]}\varphi_{free}(\tau,\xi), 
\label{eq59}
\end{equation}
where $p(\tau)$ is given by (\ref{eq57}). 

Thus, we have found the desired transformation, and it turned out to be 
non-unitary. Interestingly, it appears that this result was not previously 
known. We believe that no one has ever looked for the aforementioned 
transformation that connects a linear potential and a free particle due to 
the limited use of non-unitary transformations. 

For transparency and comparison with previous results, these transformations 
can be simplified by setting $B(t)=g=\mathrm{const}$, $k=1$, 
$c=1/\tau_0$, $d=0$, $c_1=0$ and $c_2=0$ ($\tau_0$ is some constant with the 
dimension of time). As a result, we get 
\begin{equation}
\begin{aligned}
& t=-\frac{\tau_0^2}{\tau},\;\;\;
x=\frac{\xi\tau_0}{\tau}-\frac{g\tau_0^4}{2\tau^2},\;\;\;
f_3=\frac{\xi^2}{2\tau}-\frac{g\xi\tau_0^3}{\tau^2}+
\frac{g^2\tau_0^6}{3\tau^3}, \\
& \varphi_{grv}(t,x)=\sqrt{\frac{\tau}{\tau_0}}\, 
e^{-i\left[\frac{\xi^2}{2\tau}-\frac{g\xi\tau_0^3}{\tau^2}+
\frac{g^2\tau_0^6}{3\tau^3} \right]}\varphi_{free}(\tau,\xi).
\label{eq61}
    \end{aligned}
\end{equation}
This transformation, in its simplest form, is a new equivalence between the 
time-independent linear potential and the case of free particles. 

Now consider the case $c=0$. Then $\Omega=\frac{a}{d}=\mathrm{const}$ and
without loss of generality we can assume $\Omega=1$. In this case, the equation
for $h(\tau)$ according to (\ref{eq35}) becomes
\begin{equation}
\frac{d^2h(\tau)}{d \tau^2}=-B(t(\tau)),
\label{eq62}
\end{equation}
with a solution (for simplicity, we dropped the integration constants)
\begin{equation}
h(\tau)=-\int d\tau \int B(t(\tau))\,d\tau.
\label{eq63}
\end{equation}
Correspondingly, (\ref{eq34})  will gives us $p(\tau)$:
\begin{equation}
p(\tau)=\int\left[-\frac{1}{2}\left(\int B(t(\tau))d\tau\right)^2-B(t(\tau))
\int d\tau \int  B(t(\tau))\,d\tau\right] d\tau,
\label{eq64}
\end{equation}
and we get the following transformation of coordinates, along with the 
corresponding wave function transformation: 
\begin{eqnarray} &&
\tau = t,\;\;\;x=\xi-\int d\tau \int  B(t(\tau))\,d\tau,\;\;\; 
f_3\left( \xi,\tau \right) =\xi\int B(t(\tau))\,d\tau +p(\tau), 
\nonumber \\ &&
\varphi_{grv}(t,x)= e^{-i\left[\xi\int B(t(\tau))\,d\tau +p\left( \tau \right)\right]}
\varphi_{free}(\tau,\xi).
\label{eq65}
\end{eqnarray}
Using integration by parts, we can transform the second term in (\ref{eq64})
as follows
$$\begin{aligned}
& \int -B(t(\tau))\left [\int d\tau \int  B(t(\tau))\,d\tau\right] d\tau=
\int B(t(\tau))h(\tau)\,d\tau= \\
& -\int \frac{d^2h(\tau)}{d\tau^2}h(\tau)\,d\tau=
-h(\tau)\frac{dh(\tau)}{d\tau}+\int\left(\frac{dh(\tau)}{d\tau}
\right)^2d\tau.\end{aligned}$$
Therefore,
$$p(\tau)=\frac{1}{2}\int\left(\frac{dh(\tau)}{d\tau}\right)^2d\tau-
h(\tau)\frac{dh(\tau)}{d\tau},$$
and we get
\begin{eqnarray} &&
f_3\left( \xi,\tau \right) =-\left (\xi+h(\tau)\right )\frac{dh(\tau)}{d\tau}+
\frac{1}{2}\int\left(\frac{dh(\tau)}{d\tau}\right)^2d\tau, \nonumber \\ &&
\varphi_{grv}(t,x)= e^{i\left[\left (\xi+h(\tau)\right )\frac{dh(\tau)}{d\tau}-
\frac{1}{2}\int\left(\frac{dh(\tau)}{d\tau}\right)^2d\tau\right ]}
\varphi_{free}(\tau,\xi),
\label{eq66}
\end{eqnarray}
where $h(\tau)$ is a solution of (\ref{eq61}). This is exactly the 
transformation that was presented in \cite {30} as an implementation of the 
well-known weak equivalence principle for the time-dependent gravitational 
field. It becomes much more recognizable when $B(t)=g$ is a constant. Then 
the transformation takes the form
\begin{eqnarray} &&
t=\tau,\;\;\;x=\xi -\frac {g\tau^2 }{2},\;\; u=v+f_3=v+g\xi\tau-
\frac {g^2\tau^3 }{3}, \nonumber \\ &&
\varphi_{grv}(t,x)= 
e^{-i\left[g\xi\tau-\dfrac {g^2\tau^3 }{3}\right]}\varphi_{free}(\tau,\xi).
\label{eq67}
\end{eqnarray}
This transformation is known to be a quantum formulation of Einstein's 
principle of equivalence in the context of nonrelativistic quantum mechanics 
\cite{31}.

\section{Conclusions}
In this note, we discussed the relationship between the wave functions of 
a harmonic oscillator and a free particle using the Eisenhart lift method. 
We show that only two types of potentials, namely quadratic and linear 
can be mapped to the zero potential case. This somewhat surprising result
has simple geometric interpretation: the corresponding Eisenhart manifolds
are flat.  

Only one-dimensional dynamical systems and their Eisenhart lifts have been 
considered. However, the beautiful result that the linearizability criteria 
are equivalent to the requirement that the underlying manifold be flat is 
true in the much broader context of ordinary differential equations that can 
be projected from a system of geodesic equations  \cite{32,33}. 

In this article, we did not consider the possibility of non-trivial values of 
$ g(x,t)$ in (\ref{eq11}), which will further expand the possibilities of 
including other potentials, such as the Morse potential, damped harmonic 
oscillator, etc. Research in this direction has the potential to be very 
interesting. 

In conclusion, the study of nonrelativistic physics through the prism  of 
higher-dimensional relativistic theories opens up many opportunities for 
further understanding of various aspects of nonrelativistic phenomena and 
can be useful, for example, in nonrelativistic holography \cite{21,19}, which 
has been gaining attention lately. The Eisenhart Lift Method seems to be 
a useful tool in this regard, and we hope to conduct further research in this 
direction. 

\section*{Acknowledgments}
We are grateful to  Peter Horvathy, Ole Steuernagel, Apostolos Pilafts and 
Kiyoshi Shiraishi for useful correspondence. The work is supported by the 
Ministry of Education and Science of the Russian Federation.


\begin{thebibliography}{9}
\bibitem{1A}
S.-S.~Chern, From Triangles to Manifolds,
Amer. Math. Monthly {\bf 86} (1979), 339--349. 
\url{https://doi.org/10.2307/2321093}
%%CITATION = doi:10.2307/2321093;%%

\bibitem{1B}
A.~Weinstein, Symplectic geometry, Bull. Amer. Math. Soc. {\bf 5} (1981), 
1--13.
\url{https://doi.org/10.1090/S0273-0979-1981-14911-9} 
%%CITATION = doi:10.1090/S0273-0979-1981-14911-9;%%

\bibitem{1C}
V.~I.~Arnold, {\it Mathematical Methods of Classical Mechanics} (Springer: 
New York, 1989).
\url{https://doi.org/10.1007/978-1-4757-2063-1}
%%CITATION = doi:10.1007/978-1-4757-2063-1;%%

\bibitem{1D}
H.~Geiges, A Brief History of Contact Geometry and Topology,
Expo. Math. {\bf 19} (2001), 25--53.
\url{https://doi.org/10.1016/S0723-0869(01)80014-1}
%%CITATION = doi:10.1016/S0723-0869(01)80014-1;%%

\bibitem{1E}
P.~ \v{S}evera, Contact geometry in Lagrangean mechanics,
J. Geom. Phys. {\bf 29} (1999), 235--242.
\url{https://doi.org/10.1016/S0393-0440(98)00037-0}
%%CITATION = doi:10.1016/S0393-0440(98)00037-0;%%

\bibitem{1F}
G.~B.~Halsted, Biography: Paenutij Lvovitsch Tchebychev,
Am. Math. Monthly {\bf 2N3} (1895), 61--63.
\url{https://doi.org/10.2307/2969930}
%%CITATION = doi:10.2307/2969930;%%

\bibitem{1G}
K.~Morand, Embedding Galilean and Carrollian geometries I. Gravitational waves,
J. Math. Phys. {\bf 61} (2020), 082502.
\url{https://doi.org/10.1063/1.5130907}
%%CITATION = doi:10.1063/1.5130907;%%

\bibitem{1H}
M.~Atiyah, Einstein and Geometry, Curr. Sci. {\bf 89} (2005), 2041--2044.
\url{https://doi.org/10.1142/9789812772718_0002}
%%CITATION = doi:10.1142/9789812772718_0002;%%

\bibitem{1AA}
A.~McInerney, {\it First Steps in Differential Geometry: Riemannian, Contact, 
Symplectic} (Springer: New York, 2013).
\url{https://doi.org/10.1007/978-1-4614-7732-7}
%%CITATION = doi:10.1007/978-1-4614-7732-7;%%

\bibitem{1BB}
P.~Bargue\~no,
Nonrelativistic gauged quantum mechanics: From Kaluza-Klein compactifications 
to Bargmann structures,
Phys. Lett. A {\bf 379} (2015), 1563-1567.
\url{https://doi.org/10.1016/j.physleta.2015.02.047}
%%CITATION = doi:10.1016/j.physleta.2015.02.047;%%

\bibitem{1CC}
X.~Bekaert and K.~Morand,
Embedding nonrelativistic physics inside a gravitational wave,
Phys. Rev. D {\bf 88} (2013), 063008.
\url{https://doi.org/10.1103/PhysRevD.88.063008}
%%CITATION = doi:10.1103/PhysRevD.88.063008;%%

\bibitem{1DD}
C.~Duval, G.~Burdet, H.~P.~K\"{u}nzle and M.~Perrin,
Bargmann Structures and Newton-cartan Theory,
Phys. Rev. D {\bf 31} (1985), 1841--1853.
\url{https://doi.org/10.1103/PhysRevD.31.1841}
%%CITATION = doi:10.1103/PhysRevD.31.1841;%%

\bibitem{1} 
L.~P.~Eisenhart, Dynamical trajectories and geodesics,
Annals. Math. {\bf 30} (1928-1929) 591--606.
\url{https://doi.org/10.2307/1968307}
%%CITATION = doi:10.2307/1968307;%%

\bibitem{2}
M.~W.~Brinkmann, On Riemann spaces conformal to Euclidean space,
Proc. Natl. Acad. Sci. U.S. {\bf 9} (1923) 1--3.
\url{https://doi.org/10.1073/pnas.9.1.1}
%%CITATION = doi:10.1073/pnas.9.1.1;%%

\bibitem{2A}
M.~W.~Brinkmann, Einstein spaces which are mapped conformally on each other,
Math. Ann. {\bf 94} (1925) 119--145.
\url{https://doi.org/10.1007/BF01208647}
%%CITATION = doi:10.1007/BF01208647;%%

\bibitem{3}
M.~Cariglia, C.~Duval, G.~W.~Gibbons and P.~A.~Horvathy, Eisenhart lifts and 
symmetries of time-dependent systems, Annals Phys. {\bf 373} (2016) 631--654.
\url{https://doi.org/10.1016/j.aop.2016.07.033}
%%CITATION = doi:10.1016/j.aop.2016.07.033;%%

\bibitem{3A}
M.~Cariglia, Hidden symmetries of Eisenhart-Duval lift metrics and the Dirac 
equation with flux, Phys. Rev. D {\bf 86} (2012), 084050.
\url{https://doi.org/10.1103/PhysRevD.86.084050}
%%CITATION = doi:10.1103/PhysRevD.86.084050;%%

\bibitem{3B}
G.~W.~Gibbons and C.~N.~Pope,
Kohn's Theorem, Larmor's Equivalence Principle and the Newton-Hooke Group,
Annals Phys. {\bf 326} (2011), 1760-1774.
\url{https://doi.org/10.1016/j.aop.2011.03.003}
%%CITATION = doi:10.1016/j.aop.2011.03.003;%%

\bibitem{K1}
K.~Finn, S.~Karamitsos and A.~Pilaftsis,
Quantizing the Eisenhart Lift,
Phys. Rev. D {\bf 103} (2021), 065004.
\url{https://doi.org/10.1103/PhysRevD.103.065004}
%%CITATION = doi:10.1103/PhysRevD.103.065004;%%

\bibitem{K2}
K.~Finn, S.~Karamitsos and A.~Pilaftsis,
Eisenhart lift for field theories,
Phys. Rev. D {\bf 98} (2018), 016015.
\url{https://doi.org/10.1103/PhysRevD.98.016015}
%%CITATION = doi:10.1103/PhysRevD.98.016015;%%

\bibitem{3C}
C.~Duval, M.~Hassaine and P.~A.~Horvathy,
The Geometry of Schr\"{o}dinger symmetry in gravity 
background/non-relativistic CFT, Annals Phys. {\bf 324} (2009), 1158-1167.
\url{https://doi.org/10.1016/j.aop.2009.01.006}
%%CITATION = doi:10.1016/j.aop.2009.01.006;%%

\bibitem{3D}
C.~Duval and S.~Lazzarini, Schr\"{o}dinger Manifolds,
J. Phys. A {\bf 45} (2012), 395203.
\url{https://doi.org/10.1088/1751-8113/45/39/395203}
%%CITATION = doi:10.1088/1751-8113/45/39/395203;%%

\bibitem{S1}
N.~Kan, T.~Aoyama, T.~Hasegawa and K.~Shiraishi,
Eisenhart lift for minisuperspace quantum cosmology,
arXiv:2105.09514 [gr-qc].
\url{https://arxiv.org/abs/2105.09514}
%%CITATION = ARXIV:2105.09514;%%

\bibitem{4}
M.~Cariglia, A.~Galajinsky, G.~W.~Gibbons and P.~A.~Horvathy,
Cosmological aspects of the Eisenhart-Duval lift,
Eur. Phys. J. C {\bf 78} (2018), 314.
\url{https://doi.org/10.1140/epjc/s10052-018-5789-x}
%%CITATION = doi:10.1140/epjc/s10052-018-5789-x;%%

\bibitem{4A}
E.~Minguzzi,
Eisenhart's theorem and the causal simplicity of Eisenhart's spacetime,
Class. Quant. Grav. {\bf 24} (2007), 2781--2808.
\url{https://doi.org/10.1088/0264-9381/24/11/002}
%%CITATION = doi:10.1088/0264-9381/24/11/002;%%

\bibitem{5}
M.~Cariglia and F.~K.~Alves, The Eisenhart lift: a didactical introduction of 
modern geometrical concepts from Hamiltonian dynamics, 
Eur. J. Phys. {\bf 36} (2015), 025018.
\url{https://doi.org/10.1088/0143-0807/36/2/025018}
%%CITATION = doi:10.1088/0143-0807/36/2/025018;%%

\bibitem{5A}
E.~Minguzzi,
Classical aspects of lightlike dimensional reduction,
Class. Quant. Grav. {\bf 23} (2006), 7085--7110.
\url{https://doi.org/10.1088/0264-9381/23/23/029}
%%CITATION = doi:10.1088/0264-9381/23/23/029;%%

\bibitem{5AA}
C.~W.~Misner, K.~S.~Thorne and J.~A.~Wheeler, {\it Gravitation}
(Princeton University Press: Princeton, 2017).

\bibitem{5B}
K.~F.~Sundman, M\'{e}moire sur le probl\'{e}m des trois corps, 
Acta Math. {\bf 36} (1912-1913), 105--179.
\url{https://doi.org/10.1007/BF02422379}
%%CITATION = doi:10.1007/BF02422379;%%

\bibitem{5C}
P.~Guha, B.~Khanra and A.~G.~Choudhury, On generalized Sundman transformation 
method, first integrals, symmetries and solutions of equations of 
Painlev\'{e}-Gambier type,  Nonlinear Anal. {\bf 72} (2010), 3247--3257.
\url{https://doi.org/10.1016/j.na.2009.12.004}
%%CITATION = doi:10.1016/j.na.2009.12.004;%%

\bibitem{5D}
J.~Barrow-Green, The dramatic episode of Sundman, Historia Mathematica
{\bf 37} (2010), 164--203.
\url{https://doi.org/10.1016/j.hm.2009.12.004}
%%CITATION = doi:10.1016/j.hm.2009.12.004;%%

\bibitem{6A}
A.~Lichnerowicz, {\it Th\'{e}ories  Relativistes de la Gravitation et de 
l'\'{E}lectro\-magnetisme} (Masson: Paris, 1955).

\bibitem{6B}
L.~P.~Eisenhart, {\it Riemannian Geometry} (Princeton University Press: 
Princeton, 1997).

\bibitem{6C}
A.~Garcia, F.~W.~Hehl, C.~Heinicke and A.~Macias,
The Cotton tensor in Riemannian space-times,
Class. Quant. Grav. {\bf 21} (2004), 1099--1118.
\url{https://doi.org/10.1088/0264-9381/21/4/024}
%%CITATION = doi:10.1088/0264-9381/21/4/024;%%

\bibitem{H1}
C.~Duval, P.~A.~Horvathy and L.~Palla,
Conformal Properties of Chern-Simons Vortices in External Fields,
Phys. Rev. D {\bf 50} (1994), 6658--6661.
\url{https://doi.org/10.1103/PhysRevD.50.6658}
%%CITATION = doi:10.1103/PhysRevD.50.6658;%%

\bibitem{H2}
C.~Duval, G.~W.~Gibbons and P.~Horvathy,
Celestial mechanics, conformal structures and gravitational waves,
Phys. Rev. D {\bf 43} (1991), 3907--3922.
\url{https://doi.org/10.1103/PhysRevD.43.3907}
%%CITATION = doi:10.1103/PhysRevD.43.3907;%%

\bibitem{H3}
C.~Duval, Z.~Horvath and P.~A.~Horvathy,
Vanishing of the conformal anomaly for strings in a gravitational wave,
Phys. Lett. B {\bf 313} (1993), 10--14.
\url{https://doi.org/10.1016/0370-2693(93)91183-N}
%%CITATION = doi:10.1016/0370-2693(93)91183-N;%%

\bibitem{H4}
P.~Horvathy, private communication.

\bibitem{6D}
D.~McMahon, {\it Relativity Demystified} (McGraw-Hill: New York, 2006).
\url{http://mhebooklibrary.com/doi/book/10.1036/0071455450}

\bibitem{6E}
M.~Gasperini, {\it Theory of Gravitational Interactions} (Springer-Verlag:
Berlin, 2013).
\url{https://doi.org/10.1007/978-3-319-49682-5}
%%CITATION = doi:10.1007/978-3-319-49682-5;%%

\bibitem{6F}
D.~F.~Carneiro, E.~A.~Freiras, B.~Goncalves, A.~G.~de Lima and I.~L.~Shapiro,
On useful conformal tranformations in general relativity,
Grav. Cosmol. {\bf 10} (2004), 305--312.
\url{https://arxiv.org/abs/gr-qc/0412113}
%%CITATION = GR-QC/0412113;%%

\bibitem{6G}
V.~Faraoni, E.~Gunzig and P.~Nardone,
Conformal transformations in classical gravitational theories and in 
cosmology, Fund. Cosmic Phys. {\bf 20} (1999), 121.
\url{https://arxiv.org/abs/gr-qc/9811047}
%%CITATION = GR-QC/9811047;%%

\bibitem{6AA}
V.~Ovsienko and S.~Tabachnikov, What is .. the Schwarzian derivative?  
Notices Am. Math. Soc. {\bf 56(1)} (2009), 34--36.
\url{https://www.ams.org/notices/200901/tx090100034p.pdf}

\bibitem{6BB}
O.~Lehto,  {\it Univalent Functions and Teichm\"{u}ller Spaces} 
(Springer: New York, 1987). 
\url{https://doi.org/10.1007/978-1-4613-8652-0}
%%CITATION = doi:10.1007/978-1-4613-8652-0;%%

\bibitem{6CC}
B.~Osgood, Old and New on the Schwarzian Derivative, in P.~Duren, J.~Heinonen,
B.~Osgood and B.~Palka (Eds.), {\it Quasiconformal Mappings and Analysis}
(Springer: New York, 1998). 
\url{https://doi.org/10.1007/978-1-4612-0605-7_16}
%%CITATION = doi:10.1007/978-1-4612-0605-7_16;%%

\bibitem{21}
F.~L.~Lin and S.~Y.~Wu,
Non-relativistic Holography and Singular Black Hole,
Phys. Lett. B {\bf 679} (2009), 65--72.
\url{https://doi.org/10.1016/j.physletb.2009.07.002}
%%CITATION = doi:10.1016/j.physletb.2009.07.002;%%
 
\bibitem{22}
R.~Penrose,
A Remarkable property of plane waves in general relativity,
Rev.\ Mod.\ Phys.\  {\bf 37} (1965), 215--220.
\url{https://doi.org/10.1103/RevModPhys.37.215}
%%CITATION = doi:10.1103/RevModPhys.37.215;%%

\bibitem{23}
M.~Blau, M.~Borunda, M.~O'Loughlin and G.~Papadopoulos,
Penrose limits and space-time singularities,
Class.\ Quant.\ Grav.\  {\bf 21} (2004), L43--L49.
\url{https://doi.org/10.1088/0264-9381/21/7/L02}
%%CITATION = doi:10.1088/0264-9381/21/7/L02;%%

\bibitem{23A}
A.~Yariv, {\it Quantum Electronics} (Wiley: New York, 1967).

\bibitem{23B}
O.~Steuernagel, Equivalence between free quantum particles and those in 
harmonic potentials and its application to instantaneous changes,
Eur. Phys. J. Plus {\bf 129} (2014), 114.
\url{https://doi.org/10.1140/epjp/i2014-14114-3}
%%CITATION = doi:10.1140/epjp/i2014-14114-3;%%

\bibitem{23C}
O.~Steuernagel, Equivalence between focused paraxial beams and the quantum 
harmonic oscillator, Am. J. Phys. {\bf 73} (2005), 625--629.
\url{https://doi.org/10.1119/1.1900099}
%%CITATION = doi:10.1119/1.1900099;%%

\bibitem{23D}
U.~Niederer,
The maximal kinematical invariance group of the harmonic oscillator,
Helv. Phys. Acta {\bf 46} (1973), 191--200.
\url{https://doi.org/10.5169/seals-114478}
%%CITATION = doi:10.5169/seals-114478;%%

\bibitem{8}
S.~Takagi, Equivalence of a Harmonic Oscillator to a Free Particle,
Prog. Theor. Phys.  {\bf 84} (1990), 1019--1024. 
\url{https://doi.org/10.1143/ptp/84.6.1019}
%%CITATION = doi:10.1143/ptp/84.6.1019;%%

\bibitem{O1}
M.~Oliva, D.~Kakofengitis and O.~Steuernagel,
Anharmonic quantum mechanical systems do not feature phase space trajectories,
Physica A {\bf 502} (2018) 201--210.
\url{https://doi.org/10.1016/j.physa.2017.10.047}
%%CITATION = doi:10.1016/j.physa.2017.10.047;%%

\bibitem{O2}
D.~Kakofengitis, M.~Oliva and O.~Steuernagel,
Wigner's representation of quantum mechanics in integral form and its 
applications, Phys. Rev. A {\bf 95} (2017), 022127.
\url{https://doi.org/10.1103/PhysRevA.95.022127}
%%CITATION = doi:10.1103/PhysRevA.95.022127;%%

\bibitem{6DD}
R.~L.~Devaney and L.~Keen, Dynamics of meromorphic  maps with polynomial  
Schwarzian derivative,  Ann. Sci. \'{E}cole Norm. Sup. {\bf 22}  (1989),
55--81.

\bibitem{26}
V.~Aldaya, F.~Coss\'{i}o, J.~Guerrero and F.~F.~L\'{o}pez-Ruiz,
The quantum Arnold transformation, J. Phys. A: Math. Theor. {\bf 44} (2011), 
065302. 
\url{https://doi.org/10.1088/1751-8113/44/6/065302}
%%CITATION = doi:10.1088/1751-8113/44/6/065302;%%

\bibitem{25}
O.~Steuernagel,  Equivalence between free quantum particles and those in 
harmonic potentials and its application to instantaneous changes,
Eur. Phys. J. Plus {\bf 129} (2014), 114. 
\url{https://doi.org/10.1140/epjp/i2014-14114-3}
%%CITATION = doi:10.1140/epjp/i2014-14114-3;%%

\bibitem{27}
K.~Andrzejewski and S.~Prencel,
Niederer's transformation, time-dependent oscillators and polarized 
gravitational waves,  Clas. Quantun Grav. {\bf 36} (2019), 155008.
\url{https://doi.org/10.1088/1361-6382/ab2394}
%%CITATION = doi:10.1088/1361-6382/ab2394;%%

\bibitem{27A}
Q.~L.~Zhao, P.~M.~Zhang and P.~A.~Horvathy,
Time-dependent conformal transformations and the propagator for quadratic 
systems, arXiv:2105.07374 [quant-ph].
\url{https://arxiv.org/abs/2105.07374}
%%CITATION = ARXIV:2105.07374;%%

\bibitem{29}
R.~Colella and A.~W.~Overhauser, How the COW happened, 
Physica B {\bf 385-386} (2006), 1408-1410. 
\url{https://doi.org/10.1016/j.physb.2006.05.200}
%%CITATION = doi:10.1016/j.physb.2006.05.200;%%

\bibitem{29A}
R.~Colella, A.~W.~Overhauser and S.~A.~Werner,
Observation of gravitationally induced quantum interference,
Phys. Rev. Lett.  {\bf 34} (1975), 1472-1474. 
\url{https://doi.org/10.1103/PhysRevLett.34.1472}
%%CITATION = doi:10.1103/PhysRevLett.34.1472;%%

\bibitem{29B}
U.~Bonse and T.~Wroblewski, Measurement of Neutron Quantum Interference in
Noninertial Frames, Phys. Rev. Lett. {\bf 51} (1983), 1401-1404.
\url{https://doi.org/10.1103/physrevlett.51.1401}
%%CITATION = doi:10.1103/physrevlett.51.1401;%%

\bibitem{28}
J.~J.~Sakurai, {\it Modern Quantum Mechanics} (Addison-Wesley: Reading, 1994).

\bibitem{30}
D.~Giulini,
Equivalence principle, quantum mechanics, and atom-interferometric tests.
In F.~Finster et al. (editors), {\it Quantum Field Theory and Gravity}, 
(Springer Verlag: Basel, 2012),  pp. 345-370.
\url{https://arxiv.org/abs/1105.0749}
%%CITATION = doi:10.1007/978-3-0348-0043-3_16;%%

\bibitem{31}
M.~Nauenberg, 
Einstein's equivalence principle in quantum mechanics revisited,
Am. J. Phys. {\bf 84} (2016), 879-882.
\url{https://doi.org/10.1119/1.4962981}
%%CITATION = doi:10.1119/1.4962981;%%

\bibitem{32}
A.~V.~Aminova and N.~A.-M.~Aminov,
Projective geometry of systems of second-order differential equations,
Sb. Math. {\bf 197} (2006), 951-975.
\url{https://doi.org/10.1070/SM2006v197n07ABEH003784}
%%CITATION = doi:10.1070/SM2006v197n07ABEH003784;%%

\bibitem{33}
A.~Qadir, 
Linearization: Geometric, Complex, and Conditional, 
J. Appl. Math., vol. {\bf 2012} (2012), 303960. 
\url{https://doi.org/10.1155/2012/303960}
%%CITATION = doi:10.1155/2012/303960;%%

\bibitem{19}
M.~Cariglia,
General theory of Galilean gravity,
Phys. Rev. D {\bf 98} (2018), 084057.
\url{https://doi.org/10.1103/PhysRevD.98.084057}
%%CITATION = doi:10.1103/PhysRevD.98.084057;%%

\end{thebibliography}
\end{document}